\documentclass[journal]{IEEEtran}

\usepackage{xcolor,soul,framed} 

\colorlet{shadecolor}{yellow}
\usepackage[pdftex]{graphicx}
\graphicspath{{../pdf/}{../jpeg/}}
\DeclareGraphicsExtensions{.pdf,.jpeg,.png}

\usepackage[cmex10]{amsmath}
\usepackage{array}
\usepackage{mdwmath}
\usepackage{mdwtab}
\usepackage{eqparbox}
\usepackage{url}
\usepackage{algorithmic}
\usepackage{algorithm}

\begin{document}
\bstctlcite{IEEEexample:BSTcontrol}
    \title{Using conservative voltage reduction and dynamic thermal rating for congestion management of power network}
  \author{Ramin~Nourollahi,~Rasoul Esmaeilzadeh, \textit{Member, IEEE}  \thanks{R. Nourollahi, is with the Faculty of Electrical and Computer Engineering, University of Tabriz, Tabriz, Iran. }
  \thanks{R. Esmaeilzadeh is with the Azarbaijan regional electric company.}
 }

\markboth{IEEE systems}{Roberg \MakeLowercase{\textit{et al.}}: High-Efficiency Diode and Transistor Rectifiers}

\maketitle

\begin{abstract}
Increasing the amount of electric power that is used on the demand side has brought more attention to the peak-load management of the distribution network (DN). The creation of infrastructures for smart grids, the efficient utilisation of the distributed network's components, and the appropriate administration of the distributed network would result in a valuable solution for the operators of the distributed network. As a result, a framework for peak-load management is given in this research. Within this framework, the real-time rating of the components and the voltage-dependent characteristics of the electric loads work together to assist the DN operator in effectively navigating peak periods. The combination of the conservation voltage reduction (CVR) and the dynamic thermal rating (DTR) of the components that make up the DN produces outcomes that are more helpful than any of these factors alone could provide. This is true even though each of these factors contributes to the efficient functioning of the DN. According to the findings, as compared to the individual implementation of CVR, the simultaneous utilisation of DTR and CVR results in a cost-savings rise at peak events that is 58.75 percentage points more than the individual implementation. In addition, a discussion is offered concerning the current difficulties that are being experienced by the feeders that are providing the voltage-dependent constant-power loads during the utilisation of the CVR, which are handled by the dynamic rating of the components that make up the DN.
  
\end{abstract}


\section{Introduction}\label{depMultimarket_intro}

\IEEEPARstart{T}{he} significant expansion of electric cars, distributed generation, and urban development necessitates the establishment of communication and control infrastructures to effectively manage the distribution networks (DNs) \cite{8862871}. Conversely, the yearly rise in load demands and the phenomenon of global warming pose a significant risk to the distribution networks (DNs) and their associated equipment, particularly during hot summer days (6678820). Three primary challenges to improving equipment capacity, whether via replacement or addition, are the privatisation of distribution businesses, financial constraints, and excessive prices. The present challenge has compelled the owners of DNs to contemplate the optimisation of their current infrastructures, perhaps necessitating the use of peak management strategies by the DN operator during periods of heavy loading.  The establishment of communication and control infrastructures presents an advantageous prospect for operators to use the sophisticated DN management system in order to efficiently oversee the network during periods of heavy usage. The comprehensive peak management system has a range of components, with particular emphasis placed on demand response programmes (DRPs) and voltage control programmes, which are deemed significant. In addition, the transformer and line facilities of the DN are often constrained by their temperature limitations in order to mitigate the risk of overheating incidents. Alongside the constraint of line capacity, the thermal constraints might impose some restrictions on the voltage control capabilities of distribution networks (DNs). Additionally, the implementation of real-time monitoring of the system's state may significantly contribute to the effective management of the DN's operation, particularly during critical periods \cite{8170257}. Therefore, the use of dynamic thermal rating (DTR) as opposed to traditional rating systems has potential for demand response (DN) in optimising energy management at elevated load levels, particularly during peak periods.\par
\subsection{literature review}

Voltage regulation in distribution networks (DN) refers to a control approach used to ensure that the voltage inside the DN remains within a certain range. The function of voltage regulation in the distribution network (DN) varies depending on the temporary and long-term circumstances of the network. This regulation involves adjusting the voltage levels, either by raising or lowering them, to fulfil certain objectives within the DN \cite{9444868}. Voltage regulation may be carried out in order to regulate reactive power or to avoid voltage drop in distribution network (DN) buses during periods of heavy loading. Regarding load management, voltage control may be implemented by conservative voltage reduction (CVR) in order to mitigate peak load levels and achieve energy savings \cite{Wang2014}. Experimental investigations have examined the significant impact of Conservation Voltage Reduction (CVR) on the reduction of energy consumption and demand levels in Distribution Networks (DN) \cite{Warnock1986}. The assessment of the economic and engineering advantages of the CVR in the DN's projects has been conducted in a study by \cite{Fletcher2002} alongside its technical merits. Furthermore, the paper by \cite{Shukla2018} presents a cost-benefit analysis of CVR, specifically focusing on the investment return derived from implementing CVR on distribution feeders. Furthermore, the advantages of Capacitor Voltage Regulation (CVR) have been substantiated in the context of unfavorable power system issues, such as the mitigation of power loss as examined in the study conducted by \cite{Mahendru2019}. The aforementioned advantages of CVR mostly pertain to the decrease in system demand and energy consumption resulting from voltage lowering. Another potential use of the CVR technique is in the decrease of peak loads in systems that are running in close proximity to their thermal limits \cite{Faruqui2017}. Various approaches were presented by \cite{Wang2014} to measure the impacts of the CVR (Conservation Voltage Reduction) at peak periods in the distribution network. All of the approaches used in the study demonstrated the substantial demand-saving effects of CVR at peak times, resulting in significant economic gains.
\par
Transitioning from the traditional static rating approach to the real-time dynamic rating technique would result in an augmentation of the nominal current capacity of distribution networks, particularly in relation to transformers and other components. Multiple scholarly articles have presented thermal models that may be used for the purpose of real-time monitoring of DTR \cite{Huang2013}. The Large Electric Systems Working Group (CIGRE) was first proposed by the International Council as a means to enhance the capacity of power transmission lines in a cost-effective way, particularly in comparison to other expensive approaches such as network expansion planning \cite{pramayon2010}. The findings of a concrete study on the DTR performed by the Electric Power Research Institute are elucidated in \cite{Douglass1996}. According to the findings presented in the study conducted by Douglass et al. (1996), the use of DTR techniques has been seen to enhance the capacity of transmission lines by around 1-5\%. The DTR  is influenced by several weather-related factors, such as wind speed and direction, ambient temperature, and sun irradiation. In the study by \cite{Shaker2012}, the uncertainties associated with these parameters are addressed via the use of a fuzzy-based technique. Furthermore, \cite{Zhan2017} employs three variants of polynomial regressions in their study to develop a time-series-based approach for forecasting the DTR  values of electricity lines. The DTR significantly impacts the reliability analysis of power lines.
\par
In addition, the DRP incorporates the necessary programmes for managing peak loads in the distribution network \cite{5930335}. DN's resilience, profitability, and dependability are all increased as a result of the DRP programmes' capacity to reduce the amount of power used at crucial periods \cite{20Nourollahi2021}. According to reference \cite{22Yu2018}, the network operator has a safe and efficient resource at their disposal in the form of the load of the customers who are already connected to the network in order to control the peak load.  Grid operator involvement in the smart grid residential customer's load to critical peak load management utilising the customer engagement plans is explored in the paper cited above \cite{23UlHassan2015}.  According to quote 24Alami2010, the incentive-based interruptible/curtailable DRPs are regarded to peak-load management, in which the practise of penalising consumers is considered in the event that they do not react to load reduction. This is done so that peak load management may be performed effectively.
 \par
Moreover, limited research has coordinated the CVR,  DTR,  and  DRP methods in the peak load management. In \cite{25Hossan2020}, DRP and CVR are coordinated into an advanced DN management system to increase DN efficiency by minimizing the cost of power consumption in the day-ahead market. Furthermore, most of the reviewed DTR literature is related to the transmission network level, while very limited research has studied the DTR in DN. In \cite{26Safdarian2015}, a methodology has been proposed for analyzing the potential advantages of the real-time-monitoring-based DTR in the DN from the reliability point of view.\par
From the documents that were looked through, it is possible to draw the conclusion that the CVR application in DN's peak load management solutions has not been explored in an adequate manner. In addition, DTR in DN has not been given the proper attention it deserves, particularly in the peak load management of DN, which has been preserved in its original state. The goal of the writers of this research is to discuss the CVR-based peak management of distribution networks while they are operating in overloaded situations. It is conceivable for older distribution networks to experience overloading of their substation transformers on hot summer days when there is a significant increase in the amount of power that is being used by the network. In these kinds of predicaments, the distribution network operator is required to implement the emergency DRPs, which results in an increase in the overall expense incurred by the distribution firm. In addition to that, this increase in consumption results in an increase in the cost of acquiring electricity from the network that is farther upstream. The CVR is built in such a manner so as to achieve the lowest possible cost while not violating any of the system indices in order to circumvent this issue and cut down on the expenses associated with both operational and DRP charges.
 \par
Thus, the contributions of this paper can be summarized as below:
\begin{itemize}
    \item This paper addresses the peak management issue through voltage regulation and possible thermal capacity of the network components that,to the best of our knowledge , have not been considered for peak management in the literature. 
    \item CVR method implementation provides the opportunity for peak management at a lower cost than NO-CVR condition.
    \item Considering the thermal rating of distribution network components and their dynamic characteristics, the CVR method with DTR standpoint releases the components’ capacity and   dramatically reduces the costs of peak management.
    \item The relevant power flow algorithm based on backward-forward sweep is introduced considering voltage reduction limits and dynamic line rating.
\end{itemize}
The remaining parts of the article are structured as described below. In Section 2, we go into more detail about the issue at hand and provide the mathematical formulation. In Section 3, we covered both the optimisation approach and the power flow that was applied in order to address the issue. Section 4 provides a validation of the suggested methodology, and Section 5 provides a representation of the eventual result.

\section{Problem formulation }
Within the scope of this article, a peak-load management framework for DN is put up for consideration. The suggested framework is comprised of three different methods: DRT, DRP, and CVR, all of which are capable of being implemented in the majority of DNs. At these crucial peak moments, the emergent DRP option of mandatory load reduction is being investigated as a possible means of reducing the overall stress on the network. The mathematical formulation of the aforementioned approaches will be presented in this portion of the article.

\subsection{Conservative voltage reduction}
It is reasonable to say that the electrical loads of all feeders in DN are dependent on the voltage level of the connecting bus, and this is true in both directions. In addition, the electrical loads on the DN have a unique reaction to the fluctuations in the network voltage. To represent the voltage dependence of DN'active and reactive loads, many functions have been presented in the investigations. These functions include the exponential model and the ZIP model. In this research, the ZIP model of voltage dependence is examined to model the active and reactive load changes against the voltage regulation. The goal of this modelling is to investigate the relationship between the two. You may see a representation of the mathematical formulation of the ZIP model down below \cite{32Khalili2020}.

\begin{IEEEeqnarray}{ll}
\label{EQ1}
{{P}_{n,h}}=P_{n,h}^{0}\left[ {{C}_{{{z}_{p}}}}{{\left( \frac{{{V}_{n,h}}}{V_{n,h}^{0}} \right)}^{2}}+{{C}_{{{i}_{p}}}}\left( \frac{{{V}_{n,h}}}{V_{n,h}^{0}} \right)+{{C}_{{{p}_{p}}}} \right]
\end{IEEEeqnarray}

\begin{IEEEeqnarray}{ll}
\label{EQ2}
{{Q}_{n,h}}=Q_{n,h}^{0}\left[ {{C}_{{{z}_{q}}}}{{\left( \frac{{{V}_{n,h}}}{V_{n,h}^{0}} \right)}^{2}}+{{C}_{{{i}_{q}}}}\left( \frac{{{V}_{n,h}}}{V_{n,h}^{0}} \right)+{{C}_{{{p}_{q}}}} \right]
\end{IEEEeqnarray}
In (\ref{EQ1}) and (\ref{EQ2}), $P_{n,h}^{0}$, $Q_{n,h}^{0}$, and $V_{n,h}^{0}$ represent the hourly active reactive and voltage of DN buses before implementation of the CVR. In contrast, $P_{n,h}^{{}}$, $Q_{n,h}^{{}}$, and $V_{n,h}^{{}}$ respectively refer to the hourly active, reactive and voltage of the bus n by voltage reduction. The impacts of voltage reduction on the active power of the DN can be determined through the parameters ${{C}_{{{z}_{p}}}}$, ${{C}_{{{i}_{p}}}}$, ${{C}_{{{p}_{p}}}}$ that are the active power constants of ZIP loads. Furthermore, the voltage-dependent behavior of the reactive loads determine through the constants ${{C}_{{{z}_{q}}}}$, ${{C}_{{{i}_{q}}}}$ and ${{C}_{{{p}_{q}}}}$. The voltage dependency of the DN’s loads leads to comprehensive changes in the distribution network, from power consumption to current changes, that will be evaluated in the next sections through power flow analysis.

\subsection{Dynamic thermal rating}
The traditional static thermal rating of the DN is what is used to establish the rating of the equipment in a way that is cautious and is based on the parameters that are considered to be the worst possible weather, such as greater temperatures and the lowest wind speed for overhead lines. In most cases, the static rating is established independently for each season. The operators of some DNs have the ability to manually adjust the equipment rating depending on their own personal expertise as well as the current weather conditions in order to make the most efficient use of the available network capacity. Utilising DTR in order to unlock and put to good use the DN capacity is one of the fundamental approaches that may be used when constructing the infrastructures of a smart grid. In other words, the DN operator is able to permit more power to be transported via the network components by making use of the DTR, which may result in less load curtailment occurring during the on-peak hours. The IEEE standards \cite{296692858}, \cite{306166928}, and \cite{31Degefa2012} are used in this study to simulate the DTR of the various components that make up the DN. In order for the DTR to function properly, the time constants of the components are necessary. The time constants of the component are determined by the capacitors, and the transient responses of the component are corresponding to the thermal resistance of the component. The time-constant definition is equal to the needed time interval for changing the temperature from the beginning temperature (temperature before step change in the load) to 63.2 percent of the final temperature level (final temperature a while after step change in the load). This time interval is equal to the final temperature a while after the step change in the load. There is a disparity in the time constants between the three elements that make up the DN: transformers, overhead lines, and subterranean cables. Transformers, overhead lines, and subterranean lines each have a thermal time constant that is about equal to 4 hours and 15 minutes, followed by 8 hours. This is true for each of the components that were described. While a shorter time constant makes it easier for components to cool down, a longer time constant makes them less sensitive to variations in the load that occur over short periods of time. Figure 1 illustrates a conventional thermal model for an underground cable based on the circumstances that have been provided. The thermal equivalent circuit of the DN components may be modelled in a manner similar to that of the subterranean cables seen in Fig. 1. After calculating the time constant of the components, the hourly rating of the components may be estimated by utilising the following formulae in such a way that the hottest-spot temperature of each component should be lower than its thermal limit (for example, ninety degrees Celsius). This ensures that the components will continue to function properly.

\begin{IEEEeqnarray}{ll}
\label{EQ3}
{{\theta }_{j}}({{t}_{I}})={{\theta }_{d,j}}+{{\theta }_{amb}}({{t}_{I}})+\sum\limits_{k}{{{\theta }_{j,k}}({{t}_{I}})}
\end{IEEEeqnarray}

\begin{IEEEeqnarray}{ll}
\label{EQ4}
{{\theta }_{j,k}}({{t}_{I}})= {{\theta }_{j,k}}({{t}_{I-1}})-&{{T}_{j,k}}{{W}_{c}}({{\theta }_{c}}({{t}_{I-1}})) {{e}^{-\frac{{{t}_{I}}-{{t}_{I-1}}}{{{\tau }_{k}}}}}
\nonumber \IEEEeqnarraynumspace \\
&+{{T}_{j,k}}{{W}_{c}}({{\theta }_{c}}({{t}_{I-1}}))
\end{IEEEeqnarray}
\par In (\ref{EQ3}), ${{\theta }_{j}}({{t}_{I}})$ refers to the temperature of the node $j$ between all nodes indicated in Fig. 1 at time ${{t}_{I}}$ . Moreover, ${{\theta }_{j,k}}$ represents the raised temperature arising from the dielectric losses in the node $j$ . Furthermore, ${{\theta }_{amb}}$ and ${{\theta }_{j,k}}$ respectively indicate the ambient temperature and temperature difference among nodes $j$ and $k$ . The nodal time constant is also indicated with ${{\tau }_{k}}$ in the above formulation. Besides, the control parameter ${{T}_{j,k}}$ is used in (\ref{EQ4}) to control the exponential equations, which subscripts $j$ and $k$ represent the considered node and the ladder circuit’s thermal loop, respectively. Finally, ${{W}_{c}}({{\theta }_{c}}({{t}_{I-1}}))$ refers to the conductor loss in conductor temperature ${{\theta }_{c}}({{t}_{I-1}})$ at time ${{t}_{I-1}}$ .

\par The computation of the hourly rating of DN's components is negatively affected by the inclusion of real-time DTR. This computation involves accessing the previous hour's data as well as the initial state thermal conditions of the components and the environmental states such as ambient temperature, wind speed, and sun radiation. Hence, in accordance with the dynamic alterations in the loading and environmental conditions of the component, the thermal resistances and capacitance of the components are modified using (1) and (2).  The computation of the updated ratings for overhead lines and transformers in the next hour is facilitated by the utilisation of IEEE standards 738 \cite{296692858} and IEEE standard C57.91-2011 \cite{306166928}. These standards take into account various factors such as the initial and previous hour thermal status, as well as environmental conditions including ambient temperature, solar irradiation, and wind.

\subsection{Objective function}
The distribution network operator seeks to manage the peak load in a way that minimum cost is incurred by applying the curtailment and the least acceptable system indices are met. Suppose a day that the network is in overloading condition for the peak hour period of ${{\Omega }_{h}}=\{{{h}_{1}},...,{{h}_{n}}\}$ known as event hours.  The total cost function $\Gamma $ over this period and its relevant constraints are given in (\ref{EQ5})-(\ref{EQ10}).

\begin{IEEEeqnarray}{ll}
\label{EQ5}
\Gamma =\sum\limits_{h\in {{\Omega }_{h}}}{\rho _{h}^{G}P_{h}^{G}+}\sum\limits_{h\in {{\Omega }_{h}}}{\sum\limits_{i\in {{\Omega }_{c}}}{\rho _{i,k}^{cur}\,{{\chi }_{i,h}}P_{i,h}^{0}}}
\end{IEEEeqnarray}

Subject to
\begin{IEEEeqnarray}{ll}
\label{EQ6}
P_{h}^{G}=\sum\limits_{n\in {{\Omega }_{L}}}{(1-{{\chi }_{n,h}}){{P}_{n,h}}}+{{S}_{base}}\sum\limits_{S\in {{\Omega }_{S}}}{{{r}_{s}}I_{s,h}^{2}}
\end{IEEEeqnarray}

\begin{IEEEeqnarray}{ll}
\label{EQ7}
\left[ {{\mathbf{V}}_{\mathbf{h}}}\mathbf{,}{{\mathbf{I}}_{\mathbf{h}}} \right]\mathbf{=f(\chi ,P}_{\mathbf{h}}^{\mathbf{0}}\mathbf{,Q}_{\mathbf{h}}^{\mathbf{0}}\mathbf{,Z,}V_{h}^{Sub}\mathbf{)}
\end{IEEEeqnarray}

\begin{IEEEeqnarray}{ll}
\label{EQ8}
{{V}^{\min }}\le {{V}_{n,h}}\le {{V}^{\max }}
\end{IEEEeqnarray}

\begin{IEEEeqnarray}{ll}
\label{EQ9}
{{I}_{s,h}}\times {{I}_{base}}<I_{s}^{R}{{u}_{s}}(\theta ,w,\varphi ,h)
I_{h}^{Tr}<I_{s}^{Tr,R}{{u}_{Tr}}(\theta ,w,\varphi ,h)
\end{IEEEeqnarray}

\begin{IEEEeqnarray}{ll}
\label{EQ10}
{{\chi }_{i,h}}\le MCL
\end{IEEEeqnarray}
In (\ref{EQ5}), $\rho _{h}^{G}$is the hourly market energy price and $P_{h}^{G}$ is the hourly purchased power from the upstream grid. Set of ${{\Omega }_{h}}$ denotes the set of load points that are participated in the curtailment program, $\rho _{i,k}^{cur}$ is the penalty price for every kW of load curtailment, considering the type of customers e.g. residential or industrial. Variable ${{\chi }_{i,h}}$ determines the curtailment amount of each load and $P_{i,h}^{0}$is the hourly baseline load. Equation (\ref{EQ6}) states the hourly purchased power that is sum of the CVR-effected loads in the presence of curtailment and total network loss. ${{\Omega }_{L}}$and ${{\Omega }_{S}}$ are respectively the set of load points and sections, ${{S}_{base}}$is base power, ${{r}_{s}}$is the Per-unit section resistance and ${{I}_{s,h}}$ is the Per-unit current of sections. Function $f(.)$, gives the bus voltages and section currents through the power flow in substation voltage of $V_{h}^{sub}$. $\mathbf{Z}$ is the impedance vector of the distribution network. Constraints (\ref{EQ8})-(\ref{EQ9}) state the limitations on bus voltage, section current and HV/MV transformer current $I_{h}^{Tr}$. Parameters ${{V}^{\min }}$, ${{V}^{\max }}$and ${{I}_{base}}$ respectively denote the voltage bounds and base current. $I_{s}^{R}$and $I_{s}^{Tr,R}$ respectively stand for the rating current limits of feeders and HV/MV transformers. Both ${{u}_{s}}(\theta ,w,\varphi ,h)$ and ${{u}_{Tr}}(\theta ,w,\varphi ,h)$ are the functions of temperature, wind speed, solar irradiation and hour. Inequality (\ref{EQ10}) states that the curtailment amount is restricted within a maximum curtailment level (MCL) must not exceed this value.

\section{solving algorithm}
The user's text is already academic in nature. As demonstrated in the preceding section, the aforementioned peak management issue exhibits nonlinearity and is characterised by intricate power flow limitations. In order to achieve this objective, the present study suggests the utilisation of the particle swarm optimisation algorithm (PSO) as a computational meta-heuristic approach, particularly suitable for issues involving continuous variables.The fundamental principle underlying Particle Swarm Optimisation (PSO) involves the establishment of an initial population of solutions referred to as particles, which undergo repetitive movements aimed at converging towards the optimal solution. The process of convergence is achieved by the utilisation of updating equations for the velocity and position of particles, as seen in equations (\ref{EQ11}) and (\ref{EQ12}).

\begin{IEEEeqnarray}{ll}
\label{EQ11}
Ve{{l}_{n,t+1}}=\omega Ve{{l}_{n,t}}+&{{c}_{1}}{{r}_{1}}(pbes{{t}_{n,t}}-{{Y}_{n,t}})
\nonumber \IEEEeqnarraynumspace \\
&+{{c}_{2}}{{r}_{2}}(gbes{{t}_{t}}-{{Y}_{n,t}})
\end{IEEEeqnarray}

\begin{IEEEeqnarray}{ll}
\label{EQ12}
{{Y}_{n,t+1}}={{Y}_{n,t}}+Ve{{l}_{n,t+1}}
\end{IEEEeqnarray}
In the provided equations, the variables $Ve{{l}_{n,t}}$ and ${{Y}_{n,t}}$ denote the velocity and position of the n'th particle in the t'th iteration, correspondingly. The variables $pbest$ and $gbest$ denote the optimal positions discovered by individual particles and the overall best position, respectively. The symbol $\omega$ represents the inertia weight, whereas ${{r}_{1}}$ and ${{r}_{2}}$ are random numbers that follow a uniform distribution. The acceleration coefficients are denoted as $c_1$ and $c_2$. To effectively tackle the matter of peak management, the particle placements demonstrate variability contingent upon the operator's chosen method. The fundamental purpose of the no-CVR technique is to ascertain the curtailment amounts of the generated solutions. On the other hand, the CVR and DTR methods prioritise the identification of the most favourable per-unit voltage at the substation. The configuration of particles during each hour of the event can be observed in Figure 2.

\begin{figure}[!h]
\centering\includegraphics[width=85mm]{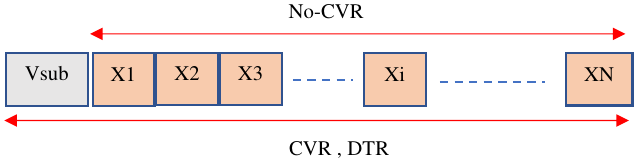}
\caption{Structure of particles for each event hour}
\label{Fig2}
\end{figure}

The CVR problem necessitates the utilisation of power flow analysis in order to determine the voltages at each bus and the currents flowing through each branch, as well as to identify any instances of voltage or current violation within the network. Due to the presence of curtailment in the loads, the use of linear power flow is rendered ineffective, resulting in the emergence of a complex nonlinear restriction. Therefore, it appears that mathematical programming optimisation models are not well-suited for addressing this particular challenge, however heuristic optimisation methods can be highly advantageous in resolving similar issues. The backward-forward sweep method is an iterative approach commonly employed for the numerical resolution of power flow problems in radial distribution networks. The power flow solution technique presented in this study is founded on the utilisation of the bus injection branch current (BIBC) matrix $\Psi$, which has a primary dimension of ${{N}_{S}}\times {{N}_{B}}$ \cite{34Teng2003}. The provided binary matrix represents the currents of each section in the network in relation to the injected currents of the load points located downstream of each section. The matrix is structured as follows. It should be noted that when constructing the BIBC matrix, it is necessary to exclude the first column in order to obtain a square matrix.

\begin{IEEEeqnarray}{ll}
\label{EQ13}
\Psi =\left( \begin{matrix}
   0 & 1 & \cdots  & 1  \\
   0 & 1 & \ddots  & 0  \\
   0 & 0 & \cdots  & 1  \\
\end{matrix} \right)
\end{IEEEeqnarray}

In order to determine the BIBC matrix of a radial distribution network, a straightforward approach is presented in the research paper by Teng et al. (2003) \cite{34Teng2003}. However, it is imperative to have a consistent numbering system for both nodes and sections, commencing with the substation and extending to the nodes further downstream. In order to address this limitation, a comprehensive algorithm is introduced in Algorithm 1, which facilitates the generation of the BIBC matrix. Algorithm 1 utilises the adjacency matrix $A$ to represent the network, where the substation nodes are uniformly designated with the value 1. Algorithm 1 is founded on the identification of leaf nodes $\Xi$ and their corresponding parent nodes $s$ in the updated tree, achieved by the iterative process illustrated in Figure 3. 

\begin{figure}[!h]
\centering\includegraphics[width=90mm]{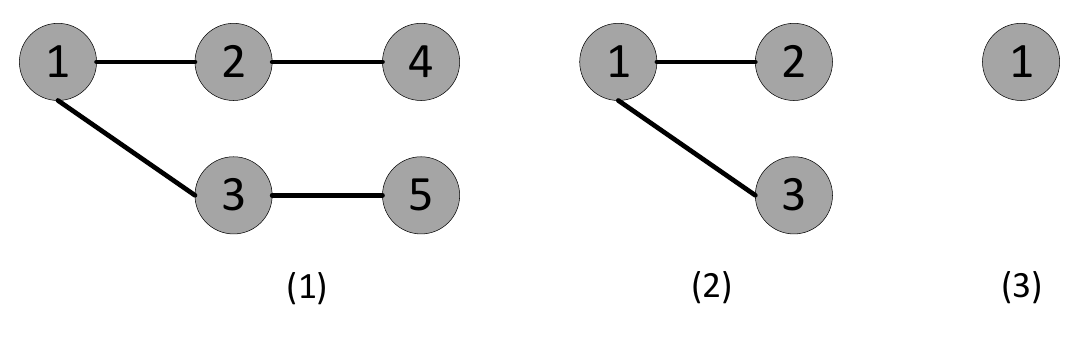}
\caption{Updated tree of BIBC matrix}
\label{Fig3}
\end{figure}

 \begin{algorithm}
 \caption{Algorithm of BIBC matrix construction}
 \begin{algorithmic}[1]
 \renewcommand{\algorithmicrequire}{\textbf{Input:}}
 \renewcommand{\algorithmicensure}{\textbf{Output:}}
 \REQUIRE \textbf{Matrix A representing network topology}
 \ENSURE \textbf{BIBC matrix} 
 \\\textit{Initialisation} :
  \STATE Construct the adjacency matrix $A$
  \STATE Create an empty cell array $\overrightarrow{\varphi }$ in length of network nodes
  \STATE $n={{N}_{S}}$
  \textit{LOOP Process}
 \WHILE {$(\Xi \ne \{\})$}
  \STATE $\Xi =\arg find(\sum\nolimits_{j}{{{A}_{i,j}}=1})$
  \FOR{$\,(i=1,i\le \left| \Xi  \right|)$}
  \STATE $r=\Xi \{i\}$
  \STATE $\phi \{r\}=[\phi \{r\},r]$
  \STATE $\Psi (n,\phi \{r\})=1$
  \STATE $s=\arg find(A(:,r)=1)$
  \STATE $A(s,r)=0,A=(r,s)=0$
  \STATE $\phi \{s\}=[\phi \{s\},\phi \{r\}]$
  \STATE $n=n-1$
  \ENDFOR
 \ENDWHILE
 \RETURN $BIBC Matrix$
 \end{algorithmic} 
 \end{algorithm}
\par Adjacent to the generation of the BIBC matrix, the backward-forward sweep can be executed using the approach outlined in approach 2. as the present iterative approach, all variables and parameters are expressed as per-unit. The technique commences by constructing a diagonal matrix ${{Z}_{D}}$, where each element corresponds to the impedance of the respective sections, as indicated in equation \eqref{EQ14}. Next, we generate a square matrix $\Upsilon$ using equation \eqref{EQ15} for utilisation in the method. The calculation of bus voltages is performed with respect to the per-unit voltage of the substation, denoted as ${{V}^{sub}}$. The functions ${{g}^{P}}$ and ${{g}^{Q}}$ defined in equations (1) and (2) are used to determine the precise values of active and reactive power for the ZIP loads in every iteration. It is important to take into account the reduced capabilities of load points when analysing the power flow through the vector variable $\chi$. Next, the current injection vector for each load is computed using the Hadamard product symbol $\odot$. After the completion of each iteration, the bus voltages are changed using the equation specified in line 8 of the method.

 \begin{algorithm}
 \caption{Algorithm of backward-forward sweep}
 \begin{algorithmic}[1]
 \renewcommand{\algorithmicrequire}{\textbf{Input:}}
 \renewcommand{\algorithmicensure}{\textbf{Output:}}
 \REQUIRE \textbf{BIBC matrix}
 \ENSURE \textbf{Results of power flow analysis (nodal voltage, branch current, loss) } 
 \\\textit{Initialisation}:
  \STATE Construct the matrix $\Upsilon $
  \STATE Create the vector of${{V}_{0}}=\overrightarrow{1}\times {{V}^{sub}}$ with length of number of load points and initiate the vector ${{\mathbf{V}}_{\mathbf{B}}}$ as the bus voltages with ${{\mathbf{V}}_{\mathbf{B}}}={{\mathbf{V}}_{0}}$
    \STATE $iter=1$
\\ \textit{LOOP Process}
  \WHILE{$(iter\le Maxiter)$}
  \STATE $\mathbf{P}={{g}^{P}}({{\mathbf{V}}_{\mathbf{B}}}\mathbf{,}{{\mathbf{C}}_{{{\mathbf{z}}_{\mathbf{p}}}}}\mathbf{,}{{\mathbf{C}}_{{{\mathbf{i}}_{\mathbf{p}}}}}\mathbf{,}{{\mathbf{C}}_{{{\mathbf{p}}_{\mathbf{p}}}}})\odot (1-\chi )\odot {{\mathbf{P}}^{\mathbf{0}}}$
  \STATE $\mathbf{Q}={{g}^{Q}}({{\mathbf{V}}_{\mathbf{B}}}\mathbf{,}{{\mathbf{C}}_{{{\mathbf{z}}_{q}}}}\mathbf{,}{{\mathbf{C}}_{{{\mathbf{i}}_{q}}}}\mathbf{,}{{\mathbf{C}}_{{{\mathbf{p}}_{q}}}})\odot (1-\chi )\odot {{\mathbf{P}}^{\mathbf{0}}}$
  \STATE $S=P+jQ,I={{\left( s\odot {{\left[ \frac{1}{V_{B}^{1}},\frac{1}{V_{B}^{2}},...,\frac{1}{V_{B}^{N}} \right]}^{T}} \right)}^{*}}$
 \STATE ${{\mathbf{V}}_{B}}={{\mathbf{V}}_{0}}-\Upsilon \mathbf{I}$
 \ENDWHILE
 \RETURN $\mathbf{V},\,\mathbf{I}$
 \end{algorithmic} 
 \end{algorithm}
 \begin{IEEEeqnarray}{ll}
\label{EQ14}
{{Z}_{D}}=\left( \begin{matrix}
   {{z}_{1}} & \cdots  & 0  \\
   \vdots  & \ddots  & \vdots   \\
   0 & \cdots  & {{z}_{N}}  \\
\end{matrix} \right)
\end{IEEEeqnarray}

\begin{IEEEeqnarray}{ll}
\label{EQ15}
\Upsilon ={{\Psi }^{T}}.{{Z}_{D}}.\Psi 
\end{IEEEeqnarray}

Figure 4 illustrates the comprehensive flowchart for addressing the issue of peak management within an overloaded distribution network. The ideal solution, which includes the substation voltage and curtailment parameters, is obtained for each event hour. In the event that the solutions produced by the Particle Swarm Optimisation (PSO) algorithm fail to adhere to the prescribed voltage and current constraints, a penalty is applied to the cost function.

\begin{figure}[!h]
\centering\includegraphics[width=90mm]{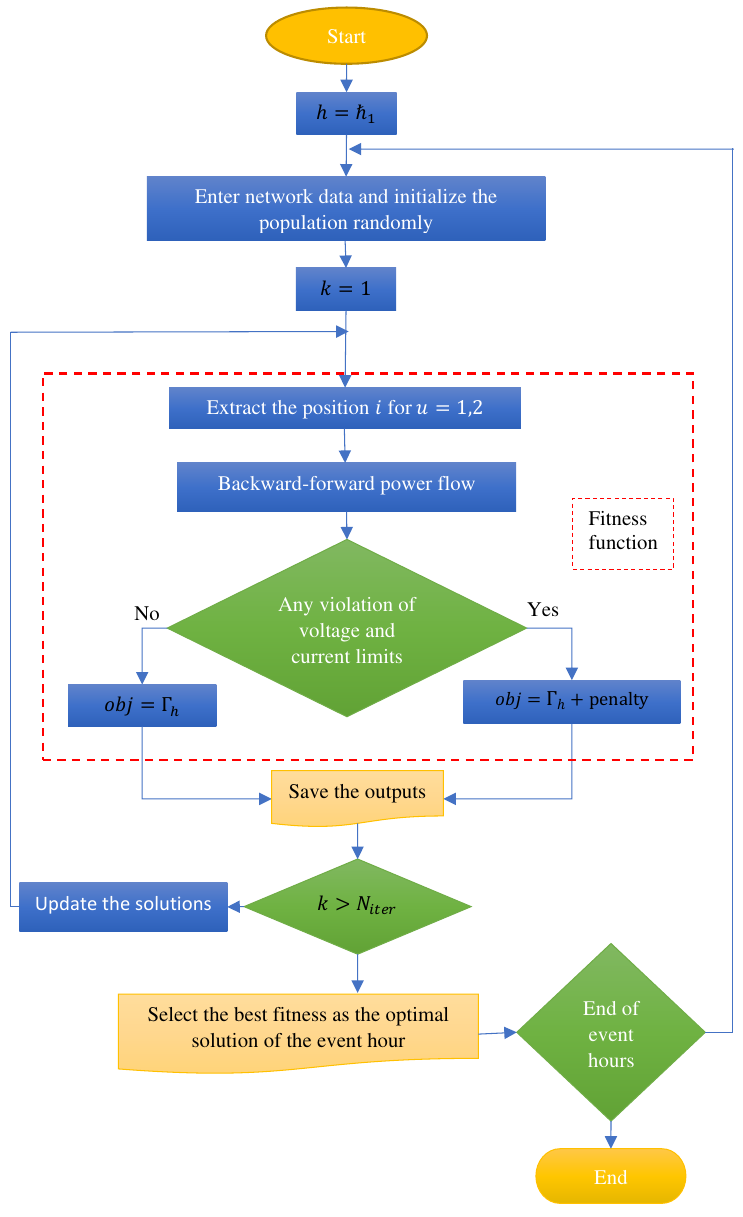}
\caption{Solving algorithm of the problem.}
\label{Fig4}
\end{figure}

\section{Simulation and Results}
\subsection{System under study}
The proposed Finnish DN bus, known as the 144 bus, is being seen as a suitable subject for examining the primary case studies discussed in this work under conditions of overload. Figure 5 depicts a concise visual representation of the system being investigated. The peak load of the test system under consideration is 11 MW per year. Additional information regarding the system being investigated, including as interruption charges, bus data, branch data, line size, and current rating, can be found in the document referenced as cite35Aaltodoc. As seen in Figure 5, the Finnish DN system for the 144 bus configuration has a single primary distribution substation operating at 110/20 kV. This substation is responsible for providing power to 144 secondary substations operating at 20/0.4 kV. The overall configuration of the system is radial in nature. In addition, the data pertaining to weather conditions, such as ambient temperature, sun irradiation, and wind speed, were sourced from the publication by Finnish et al. (2018) \cite{36Finnish2018}. The weather data pertaining to the most extreme temperature recorded throughout the summer of 2019 is utilised in the simulation procedure.

\begin{figure}[!h]
\centering\includegraphics[width=85mm]{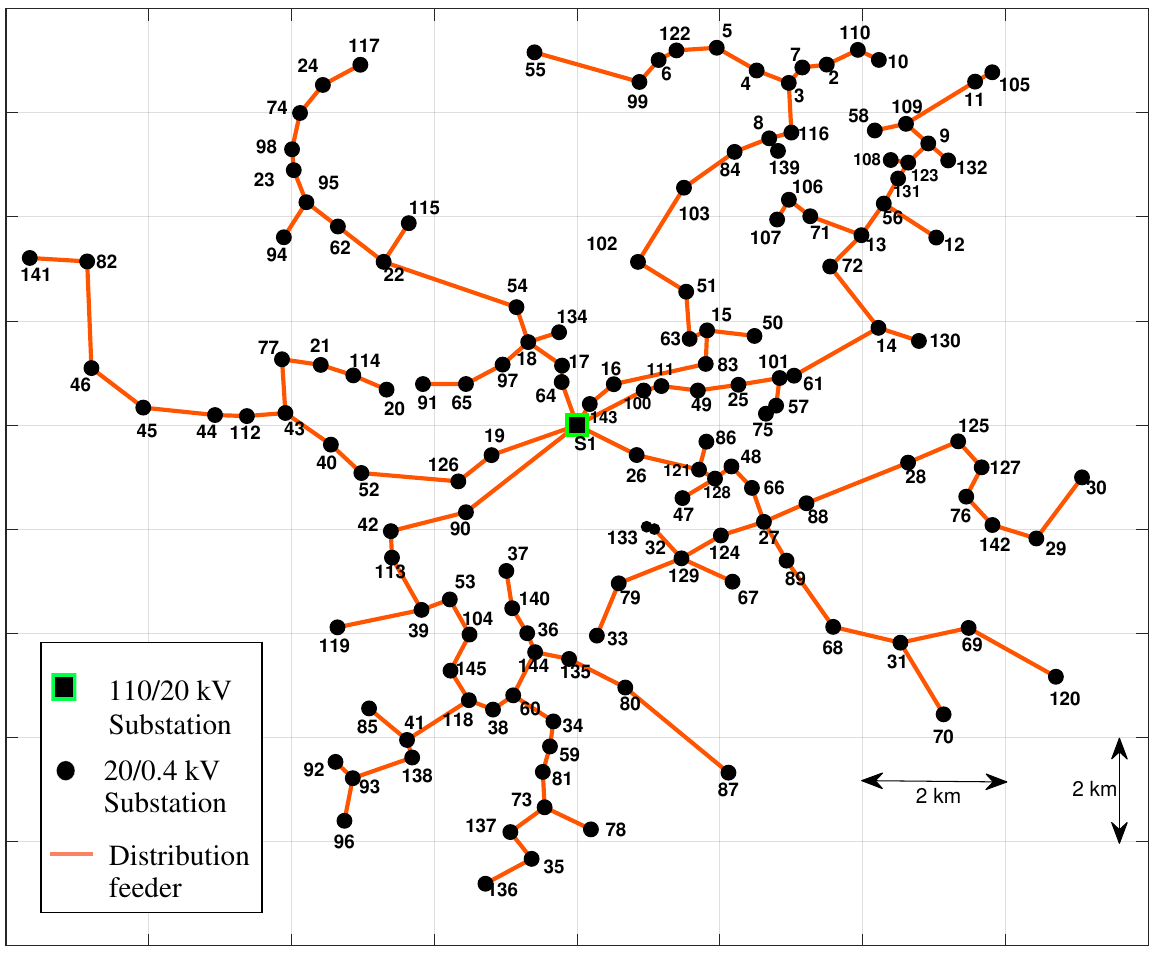}
\caption{Single line representation of the under study test system.}
\label{Fig5}
\end{figure}

\subsection{Numerical results}
The obtained results are represented in three case studies as below.
\begin{itemize}
    \item Case 1: Operation of DN in normal mode without CVR and DTR 
    \item Case 2: Operation of DN by implementation of CVR without considering DTR during the peak times (hour 10- hour 21)
    \item Case 3: Operation of DN by implementation of CVR considering DTR during the peak times (hour 10 to hour 21)
\end{itemize}
Furthermore the problems related to increased lines current by implementation of the CVR due to the constant power loads will be discussed in the following of results.
\subsubsection{Cost results}
\begin{figure}[!h]
\centering\includegraphics[width=90mm]{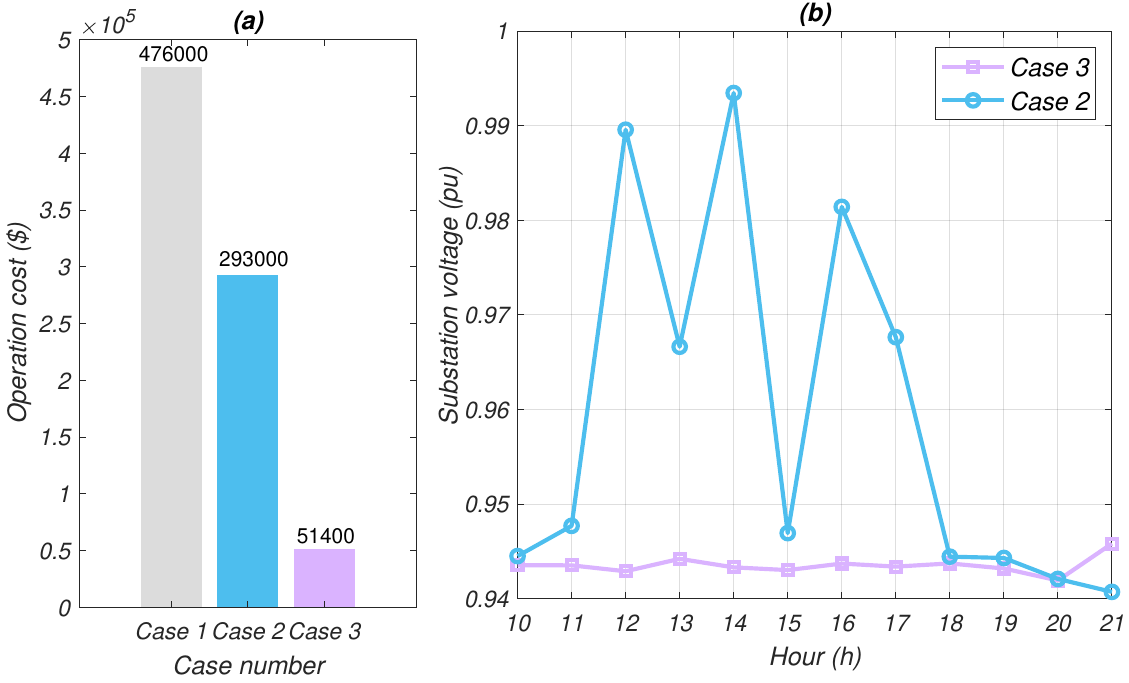}
\caption{Cost and voltage results of each case study.}
\label{Fig6}
\end{figure}

The primary benefits of the proposed CVR-DTR, integrated with the DRP framework, are evident in the reduction of operational expenses for the distribution network. The cost findings of the introduced case studies are depicted in Figure 6-(a) (see Fig. \ref{Fig6}-(a)). Based on the information shown in Figure 6-(a), it is evident that the operational cost of the DN is \$476,000, \$293,000, and \$51,400 during the peak hours (from hour 10 to hour 21) for examples 1 to 3, respectively. Based on the aforementioned findings, it can be inferred that the deployment of the CVR alone leads to a decrease of 38.45\% in system operation costs during peak times. However, this cost reduction can be further increased to 89.2\% by concurrently implementing both DTR and CVR. Furthermore, it is important to note that the decrease in cost reduction observed in case 2 compared to case 3 can be attributed to the findings depicted in Figure 6-(b), which play a crucial role in accurately interpreting the results. Based on the findings shown in Figure 6-(b), it is evident that a limitation exists which hinders further lowering of voltage in example 2. This constraint is associated with certain buses that possess substantial constant-power loads. The excessive drop in voltage at the substation might lead to increased currents flowing through the substation transformer and its associated branches, potentially exceeding their thermal current restrictions. Therefore, the ability of the substation transformer and branches to adjust their thermal rating dynamically enables a greater reduction in substation voltage in scenario 3, as depicted in Figure 6-(b) (see Fig. \ref{Fig6}-(b)). Given the significance of this matter, there has been an increased focus on the ongoing developments pertaining to the following.
 
\subsubsection{Substation power and load curtailment}
To effectively demonstrate the optimal utilisation of the substation and lines capacity in various scenarios, the conveyed power originating from the 110/20 kV substation is visually depicted in Figure 7. As previously said, the utilisation of Dynamic Voltage Regulation (DTR) during the deployment of Constant Voltage Regulation (CVR) results in increased cost reduction as a consequence of the enhanced flexibility in voltage reduction. The decrease in cost can be attributed to the less load curtailment observed in example 3 in comparison to case 2. To clarify, the adoption of Demand Turn Reduction (DTR) will result in a decrease in load curtailment during the events. Additionally, the enhanced thermal capacity of DN's components will enable the provision of electricity to a greater number of loads. This fact is depicted in Figure 7. Based on the findings presented in Figure 7, it can be inferred that the dynamic thermal rating of the distribution network's components, specifically the branches and substation transformer as discussed in this study, leads to a substantial increase in the provided load through the substation in case 3 as compared to case 2. In order to effectively showcase the effects of CVR (Conservation Voltage Reduction) and DTR (Demand Time Response) in peak duration management, we give the illustrative Figure 8, which depicts the load curtailment seen in each of the case studies. Based on the information shown in Figure 8, there is a modest decrease in curtailed loads observed in instance 2 when compared to case 1. Moreover, a notable decrease is shown as a result of the probable impacts of the DTR in scenario 3, as depicted in Figure 8.

\begin{figure}[!h]
\centering\includegraphics[width=90mm]{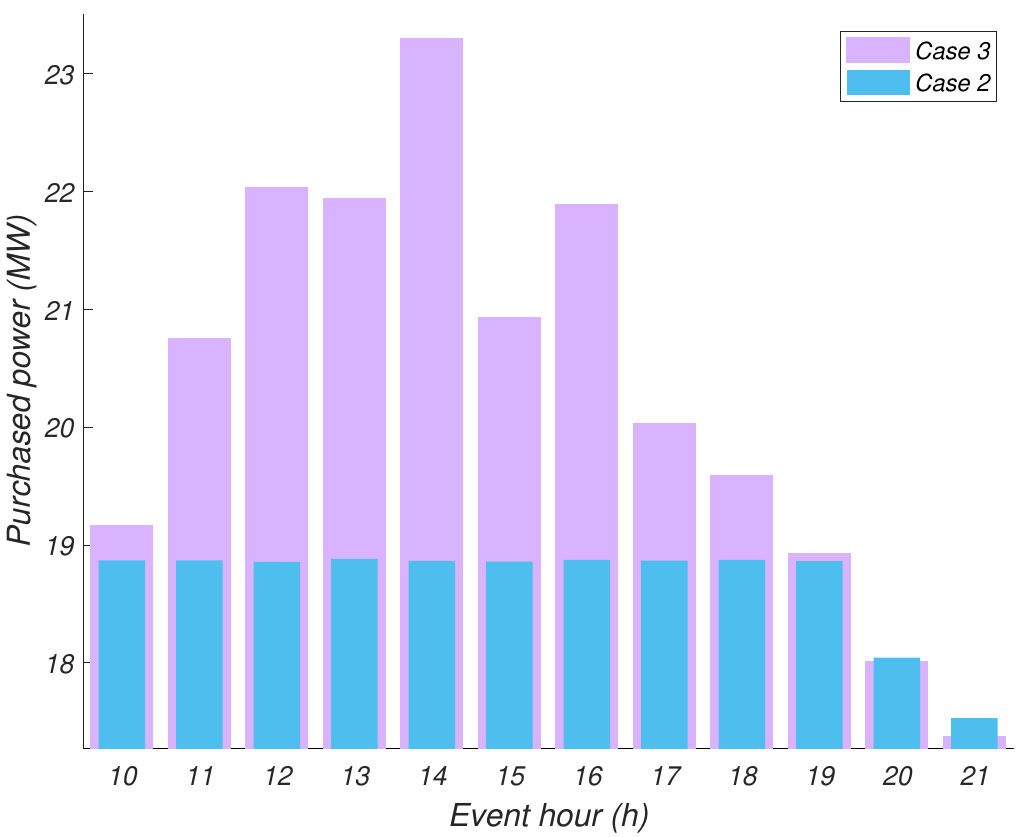}
\caption{Carried power from the 110/20 kV substation during the event.}
\label{Fig7}
\end{figure}

\begin{figure}[!h]
\centering\includegraphics[width=90mm]{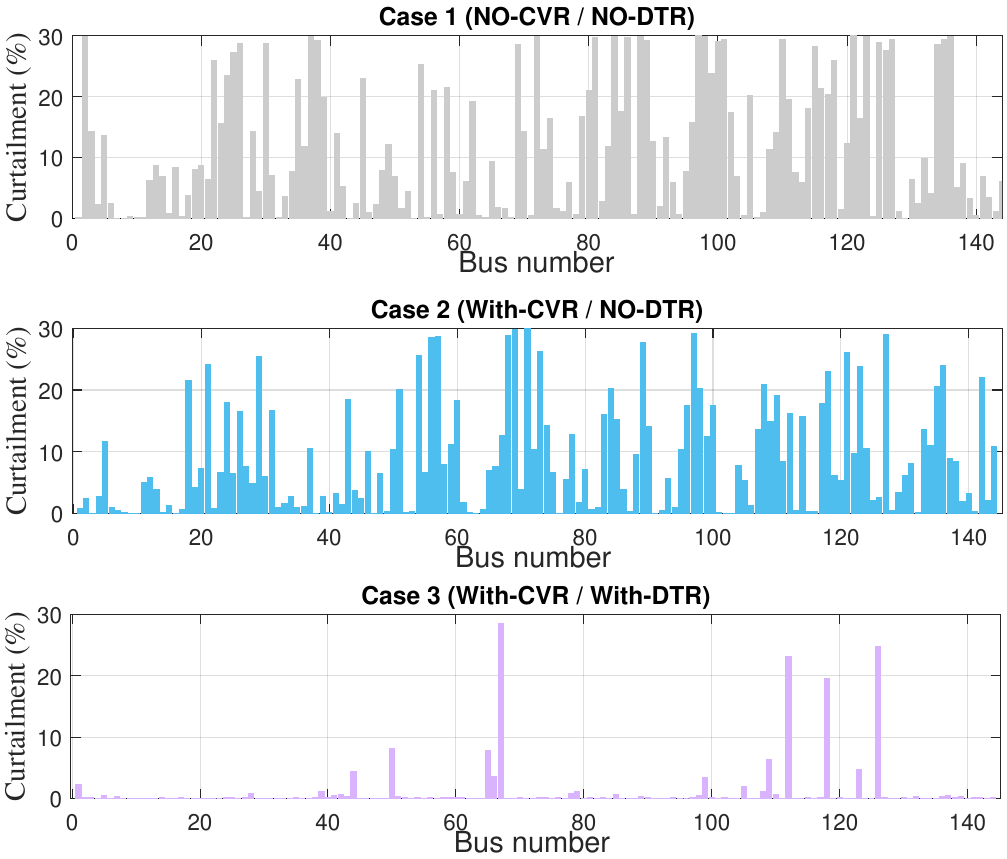}
\caption{curtailed loads of buses in each case studies.}
\label{Fig8}
\end{figure}
\subsubsection{Voltages and currents analysis}
The figure labelled as Figure 9 illustrates the voltage levels of the buses in the various case studies conducted during two specific event hours, namely 10 and 14. These case studies differ in terms of temperature, wind speed, and irradiation. Based on the information presented in Figure 9, it is apparent that the application of CVR results in a noticeable decrease in voltage in case 2 when compared to case 1. The voltage reduction observed at both 10:00 and 14:00 hours is nearly same, as depicted in Figure 9. This similarity is evident in both scenarios 1 and 2. On the other hand, in scenario 3, the varying local weather conditions (including wind, solar radiation, and temperature) between hour 10 and hour 14 contribute to an increased potential for voltage reduction in the majority of buses. Additionally, the current flowing through the branch in case studies 2 and 3 is depicted in Figure 10. Based on the information presented in Figure 10, it is evident that in Case 3, there is an increased allowance for currents to flow through the branches, particularly in the branches located in close proximity to the substation. This is mostly attributed to the potential effects of Distributed Temperature Regulation (DTR).

\begin{figure}[!h]
\centering\includegraphics[width=90mm]{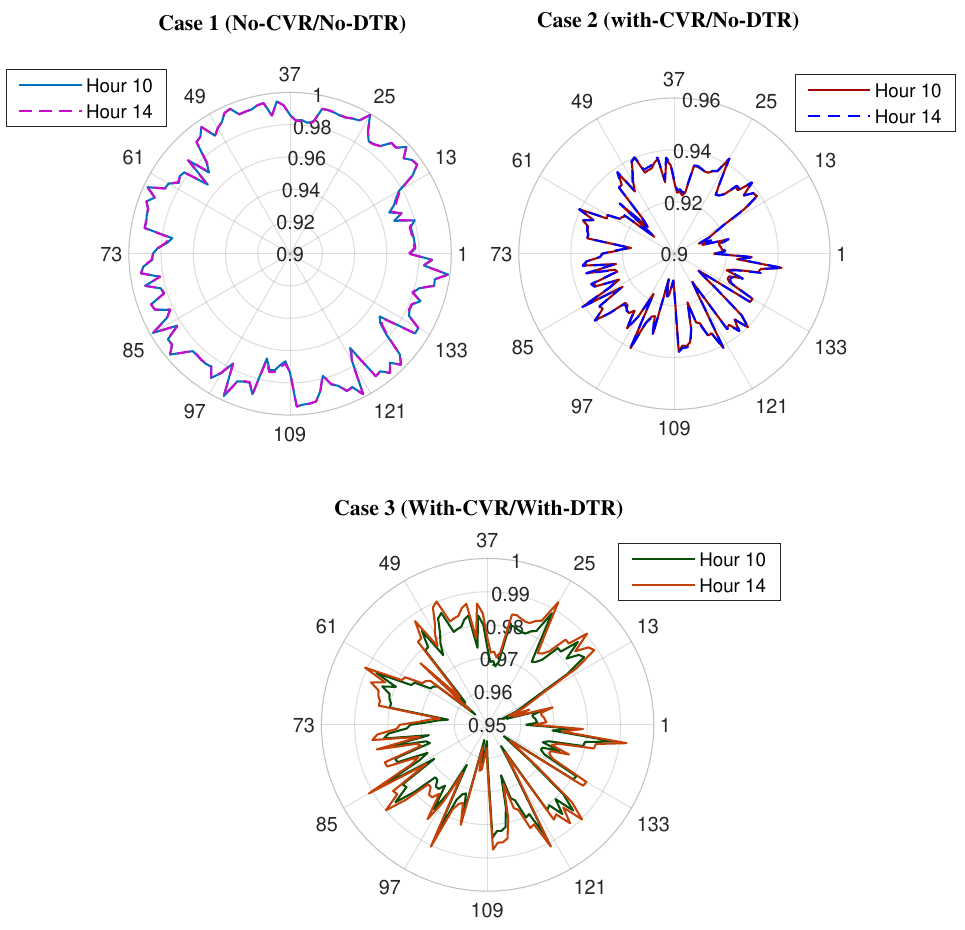}
\caption{Buses voltages in each case study.}
\label{Fig9}
\end{figure}

\begin{figure}[!h]
\centering\includegraphics[width=90mm]{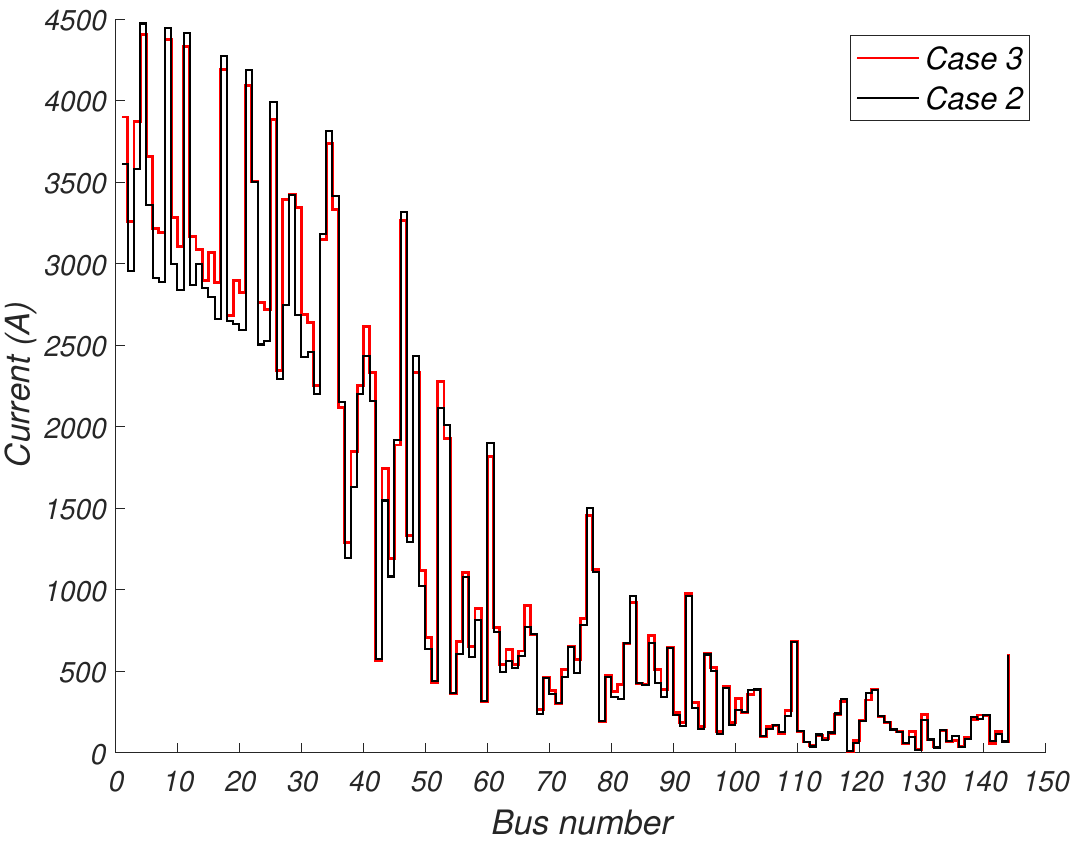}
\caption{Branches current in each case studies.}
\label{Fig10}
\end{figure}

\subsubsection{Effects of DN's demand factor on CVR potential}
As previously stated, the potential of the CVR (Conservation Voltage Reduction) in a Distribution Network (DN) is contingent upon the loading condition of the entire network, as it is constrained by the thermal limit of the DN's components. Figure 11 illustrates the potential of Demand Response (DR) in the context of Conservation Voltage Reduction (CVR) across various demand parameters. This observation is of particular significance when examining the efficacy of CVR in managing peak loads within the Distribution Network (DN). The increase in the DN demand factor leads to an observable rise in the minimum voltage of DN, as depicted in Figure 11. A considerable increase in the minimum CVR voltage is observed with a demand factor of 95\%. Therefore, it can be inferred from Figure 11 that the use of CVR-based peak load management in distribution networks (DN) can be substantially mitigated by enhancing the loading condition of the DN.

\begin{figure}[!h]
\centering\includegraphics[width=80mm]{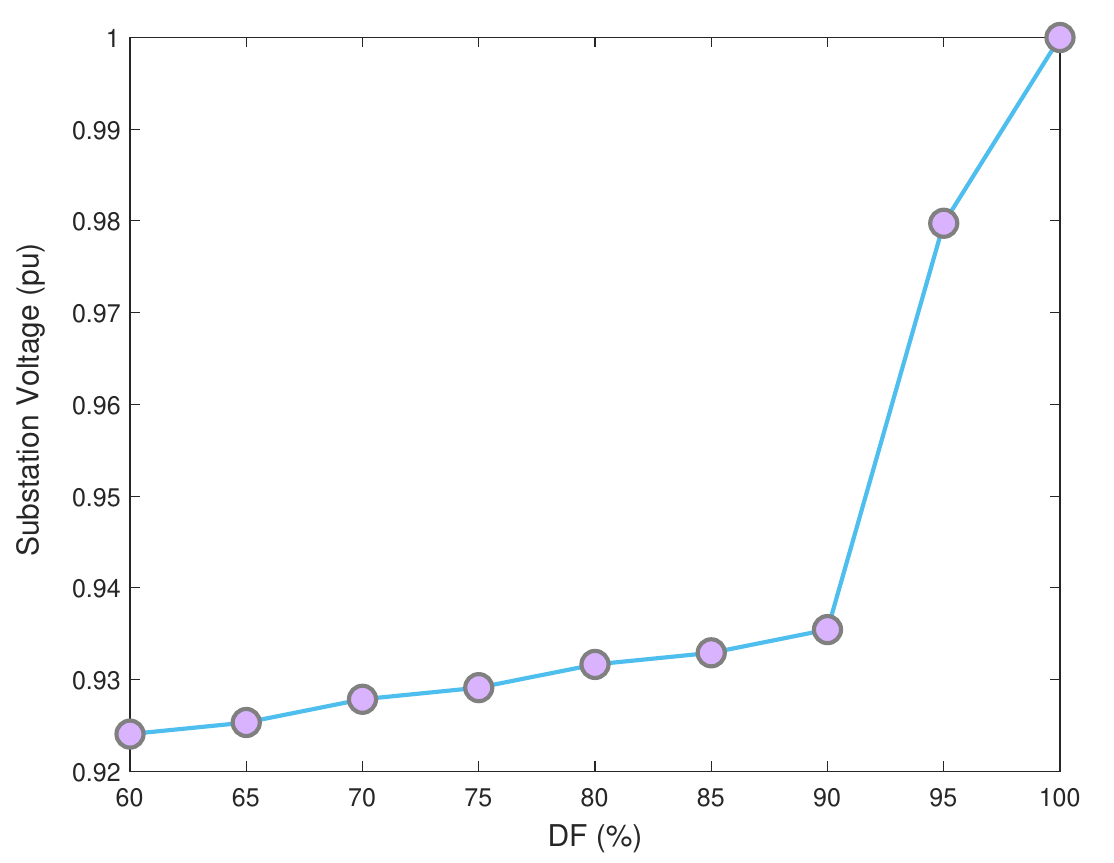}
\caption{Effects of demand factor on CVR.}
\label{Fig11}
\end{figure}
\subsubsection{Current changes of DN under CVR considering the DTR}
To effectively demonstrate the variations in currents of the feeders resulting from the adoption of CVR, a visual representation in the form of Figure 11 is utilised, employing a colour map for enhanced clarity. As anticipated, a higher magnitude of current variation is typically observed in the feeders closest to the substation. Figure 11 visually emphasises the initial feeders using warm colours, representing the more pronounced fluctuations in current inside these particular feeders. Based on the information presented in Figure 11, it can be observed that only a subset of the feeders have encountered significant fluctuations in current, which has subsequently hindered the successful implementation of CVR.

\begin{figure}[!h]
\centering\includegraphics[width=90mm]{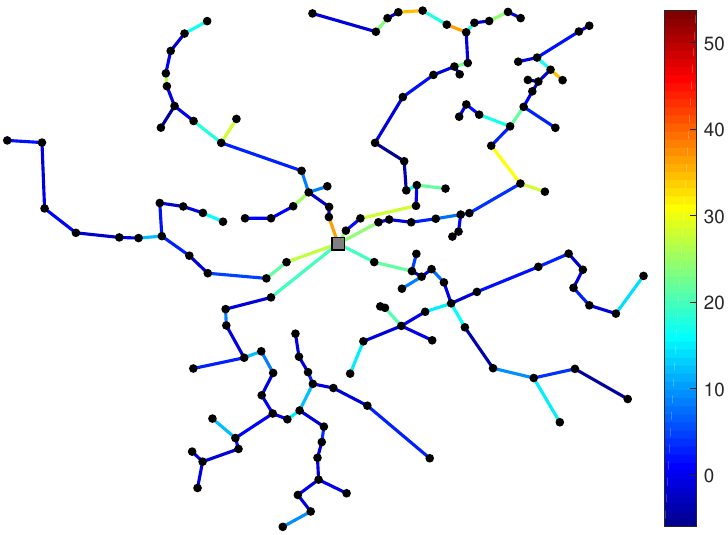}
\caption{Current change in DN's feeders.}
\label{Fig12}
\end{figure}
\vspace{-0.5cm}
\section{Conclusion}
This article investigates three peak load control strategies of the distribution network (DN), including CVR, DTR, and DRP. The CVR has been extensively examined and introduced in past scholarly works, but there has been comparatively less focus on the application of the DTR in managing peak loads in Distribution Networks . In addition to examining the direct impacts of the DTR on peak load management of the Distribution Network (DN), our investigations extend beyond prior individual studies by exploring the integration of the DTR with CVR, which offers further advantages in terms of voltage reduction potential within the DN. According to the findings presented, the utilisation of the individual CVR results in a cost reduction of 38.45\% during peak hours. In contrast, the implementation of the CVR considering DTR leads to a significantly higher cost reduction of 89.2\% within the same peak duration. Moreover, the augmentation of the feeder's current due to the implementation of the CVR has been substantiated by the findings presented in the given results. Consequently, the examination of the DTR has been undertaken as a potential resolution to address this issue. The suggested peak management framework offers the advantage of reducing load curtailment during peak events. This reduction is achieved through the simultaneous utilisation of the CVR and DTR techniques. Moreover, according to the findings, the procurement of power from the energy market experiences an increase during periods of high demand through the utilisation of the CVR and DTR strategies. It is noteworthy to acknowledge that there was a significant level of load curtailment before to the implementation of CVR and DTR. However, the introduction of DTR and CVR has effectively decreased this curtailment, resulting in the provision of power to a greater number of loads.

\end{document}